# Real-Time Quantum Dynamics of Long-Range Electronic Excitation Transfer in Plasmonic Nanoantennas

*Niranjan V. Ilawe, M. Belén Oviedo,\* and Bryan M. Wong\**

Department of Chemical & Environmental Engineering and Materials Science & Engineering Program, University of California, Riverside, Riverside, CA 92521, United States

\*Corresponding authors. E-mail: M. Belén Oviedo: boviedo@fcq.unc.edu.ar; Bryan M. Wong: bryan.wong@ucr.edu. Web: http://www.bmwong-group.com

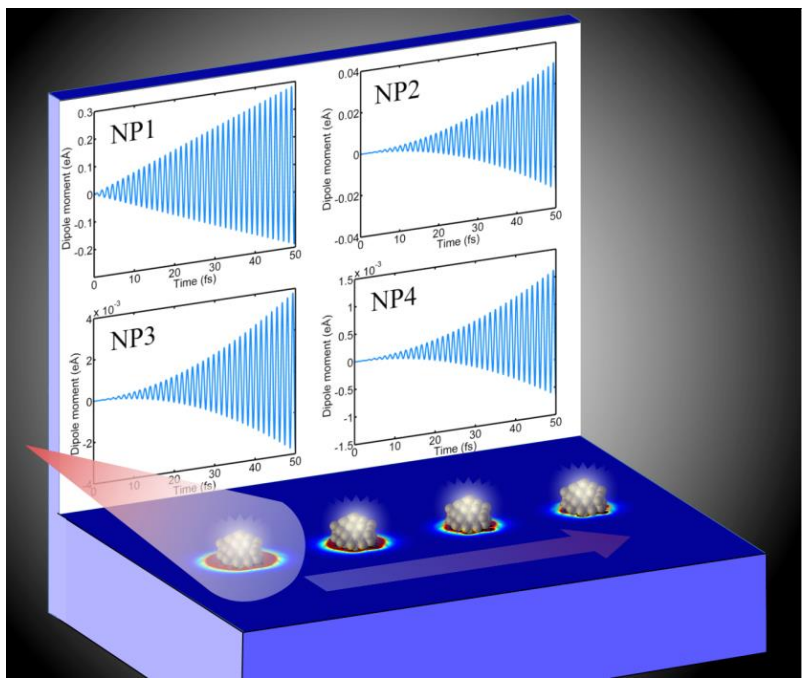

**Abstract**

Using large-scale, real-time quantum dynamics calculations, we present a detailed analysis of electronic excitation transfer (EET) mechanisms in a multi-particle plasmonic nanoantenna system. Specifically, we utilize real-time, time-dependent, density functional tight binding (RT-TDDFTB) to provide a quantum-mechanical description (at an electronic/atomistic level of detail) for characterizing and analyzing these systems, without recourse to classical approximations. We also demonstrate *highly* long-range electronic couplings in these complex systems and find that the range of these couplings is more than twice the conventional cutoff limit considered by FRET based approaches. Furthermore, we attribute these unusually long-ranged electronic couplings to the coherent oscillations of conduction electrons in plasmonic nanoparticles. This long-range nature of plasmonic interactions has important ramifications for EET – in particular, we show that the commonly used "nearest-neighbor" FRET model is inadequate for accurately characterizing EET even in simple plasmonic antenna systems. These findings provide a real-time, quantum-mechanical perspective for understanding EET mechanisms and provide guidance in enhancing plasmonic properties in artificial light-harvesting systems.

**Introduction**

The efficient harvesting of abundantly available solar energy for enhancing photochemical reactions relies on the efficient capture of photons and subsequent *transfer* of this excitation energy to the reactive site[1]. Taking inspiration from natural light-harvesting complexes, researchers have begun exploring novel plasmonic antenna systems for directing and controlling this flow of excitation energy[2]. These excitation energy transfer mechanisms are mediated by local surface plasmonic resonances[3] (LSPR) that describe the coherent oscillation of metal conduction electrons caused by the electric field of the incident photons. These LSPRs are characterized by a strong optical absorption and large electric field enhancements that are highly dependent on the nanoparticle (NP) material, size, shape, and surrounding environment[4-7]. Moreover, due to the coherent nature of these oscillating electrons, LSPRs also exhibit large dipole moments, enabling electronic excitation transfer (EET) to neighboring nanoparticles via electrostatic coupling[8]. This electrostatic coupling is analogous to Förster resonance energy transfer (FRET)[9] mechanisms seen ubiquitously in nature, and these strongly-coupled plasmonic nanoparticles have allowed several advances in plasmon-mediated excitation energy transfer processes[8, 10-12]. In particular, studies by Maier et al., have shown direct experimental evidence of EET along a plasmon waveguide made up of silver nanorods.[8] EET has also been observed in noble materials such as gold and silver nanoclusters which function as acceptors for EET[13]. Recently, Scholes and co-workers[2] have characterized plasmonic nanoantenna systems, inspired from naturally found light-harvesting systems, for use in solar fuel production.

The most widely employed approaches for analyzing EET in the previously mentioned systems are Förster's approach and classical electrodynamics theories based on solving Maxwell's equations[14-17]. Förster correlated the energy released by the de-excitation of a donor (and subsequent energy uptake by the acceptor) to the spectral overlap between the emission and absorption spectra of the donor and acceptor respectively. Förster's equation is generally applicable when the following two conditions are satisfied i) a dipolar approximation can be employed for the electronic coupling, and ii) a spectral overlap is present in the emission and absorption spectra of the donor and acceptor respectively. This theory has been successfully used for predicting EET rates in various systems such as proteins, membranes, and other biological systems[18]. The $1/R^6$ type distance dependence of EET as predicted by Förster's theory has been exploited to create chemical "rulers" to determine nanoscale distances within chemical and biological species[19-21]. A slightly modified variation of Förster's equation has also been used specifically to model the plasmon-induced EET for solar energy applications[22]. Classical electrodynamics theories based on solving Maxwell's equations, have also been used frequently to investigate the excitation transfer mechanisms in metal nanoparticle chains[14-17]. However, these models contain approximations, such as spectral overlap or the dipole approximation, which limit their applicability to more complex systems. For example, when several donor and/or acceptors are arranged in a complex or confined geometry, such as those found in photosynthetic light-harvesting antennas[2], the applicability of the classical models raises concerns[23-24]. Particularly, London and others[25-26] have revealed that for large planar NPs, the multipole expansion averages away the shape of the donor and acceptor. In such cases, it is advisable to consider local interactions between different parts of the NPs rather than the total electronic coupling as approximated by the dipole approximation. Further studies have also found that spectral overlap, as considered by Förster, neglects the contribution of optically dark states to the rate of EET[27-28]. Some of these concerns have been resolved in recent years by fascinating studies, such as transition densities obtained directly through quantum-chemical calculations[29] and variants such as distributed-monopoles[30], line-dipole approximations[31] and generalizations of Förster's theory[23]. Other studies have gone beyond the traditional dipole approximation to overcome these limitations and have applied generalized multipole techniques to include higher order multipoles in analyzing plasmon propagation along metal nanospheres,[32] nanorods,[33] and photonic crystals.[34] However, as we approach new emerging areas in mesoscale processes[35] (such as quantum coherence in biological systems and collective excitations at the nanoscale), we must re-assess the potential limitations of these simplistic models[14-17, 22-25, 36] which may be inapplicable to large, strongly-interacting, electronic systems such as plasmonic nanoantennas. Moreover, a deep understanding of the precise EET mechanisms at a quantum dynamical level of detail in these large multi-particle electronic systems is essential for guiding future experimental work to harness and control these complex systems.

In this work, we probe mechanistically the EET phenomena in large plasmonic nanoantenna systems using the density functional tight-binding (DFTB) approach[37] and its real-time time-dependent counterpart, RT-TDDFTB, without recourse to the spectral overlap or point-dipole approximations for characterizing the electronic couplings. In particular, we reveal highly long-range electronic couplings in plasmonic nanosystems that are more than twice the conventional Förster's limit considered in traditional approaches[13]. Furthermore, we show that these long-range electronic couplings not only give rise to a complex interplay of interactions between all the NPs in the nanoantenna system but also question the direct applicability of conventional theoretical models, based on classical theories, to these plasmonic nanosystems. We also propose a representative analytical model that captures the basic underlying dynamics of the full quantum dynamical method and provides a phenomenological understanding of the EET mechanism. While we focus our studies on a single representative nanoantenna system (4 icosahedral NPs, each containing 55 sodium atoms with 220 atoms total), our main qualitative results are expected to apply to a broad range of other complex plasmonic systems.

**Theory and Methodology**

The real-time electron dynamics of large plasmonic nanoantenna systems cannot be routinely probed with conventional linear-response TD-DFT or continuum models and, therefore, our use of the self-consistent density functional tight-binding (SCC-DFTB) formalism[37] is crucial to this study. The DFTB formalism is based on the second order expansion of the Kohn-Sham (KS) total energy, EKS, with respect to fluctuations of the electron density, $\rho(r) = \rho_0(r) + \delta\rho(r)$, around a reference density, $\rho_0(r)$ of neutral atomic species. The primary idea behind this method is to describe the Hamiltonian eigenstates with an atomic-like basis set and replace the Hamiltonian with a parameterized Hamiltonian matrix whose elements depend only on the internuclear distances (neglecting integrals of more than two centers) and orbital symmetries[37]. The original DFTB formulation begins with the expression for the Kohn-Sham total energy,

$$E_{\text{KS}} = \sum_i^{\text{occ}} \left\langle \psi_i \middle| \left( -\frac{1}{2}\nabla^2 + V_{\text{ext}} \right) \middle| \psi_i \right\rangle + E_{\text{H}} + E_{\text{xc}} + E_{\text{II}}, \quad (1)$$

where $\psi_i$ are the Kohn-Sham orbitals, $V_{\text{ext}}$ is the external interaction (including electron-ion interactions), $E_{\text{H}}$ is the Hartree energy, $E_{\text{xc}}$ is the exchange-correlation energy, and $E_{\text{II}}$ is the ion-ion interaction energy. On expanding the Kohn-Sham total energy in terms of a reference density and a small correction $\rho_0 + \delta\rho$, the DFTB energy is parameterized as

$$E_{\text{DFTB}} = \sum_i^{\text{occ}} \langle \phi_i | \hat{H}_0 | \phi_i \rangle + \frac{1}{2} \sum_{AB}^M \gamma_{AB} \Delta q_A \Delta q_B + E_{\text{rep}}^{AB}. \quad (2)$$

The first term in Eq. (2) corresponds to a Kohn-Sham effective Hamiltonian, $\hat{H}_0$, evaluated at the reference density $\rho_0$, and is approximated in the DFTB framework as

$$\hat{H}_0 \approx \langle \phi_\mu | \hat{T} + v_{\text{eff}}[\rho_A^0 + \rho_B^0] | \phi_\nu \rangle, \qquad \mu \in A, \nu \in B, \tag{3}$$

where $\phi_\mu$ forms a minimal Slater-type orbital basis centered on the atomic sites, $\hat{T}$ is the kinetic energy operator, $\rho_A^0$ is the reference density of the neutral atom A, and $v_{\text{eff}}$ is the effective Kohn-Sham potential. It should be noted that the Hamiltonian matrix elements depend only on atoms A and B, therefore, only two-center Hamiltonian matrix elements, as well as two-center elements of the overlap matrix, are explicitly calculated using analytical functions as per the LCAO formalism. These matrix elements are pre-tabulated for all pairs of chemical elements,[23] as a function of distance between atomic pairs, significantly improving the computational efficiency of the DFTB approach. The second term in Eq. (2) is the energy due to charge fluctuations and is parameterized analytically as a function of orbital charges and $\gamma_{AB}$, which is a function of inter-atomic separation and Hubbard parameter, $U$.[24] The quantity $\Delta q_A = q_A - q_A^0$ is the difference between the charge of the isolated atom $q_A^0$ and the charge $q_A$ obtained via a Mulliken population analysis of atom A in the molecule. $E_{\text{rep}}$ is the distance-dependent diatomic repulsive potential and contains the core electron effects, ion-ion repulsion terms as well as some exchange-correlation effects. $E_{\text{rep}}$ can be considered as a practical equivalent to an xc-functional in DFT which is approximated with simple functions in the DFTB formalism. As per the consideration of free atoms, $\rho_0$ is spherically symmetric; hence, the ion-ion repulsion can be approximated to depend only on the elements and their distance. Contributions of three and more centers are rather small and can be neglected. These pair-wise repulsive functions are obtained by fitting to DFT calculations using a suitable reference structure.[22] Assuming tightly bound electrons, and using a minimal local basis (only one radial function for each angular momentum state), the DFTB Hamiltonian is given by

$$\hat{H}_{\text{DFTB}} = \langle \phi_\mu | \hat{H}_o | \phi_\nu \rangle + \frac{1}{2} \hat{S}_{\mu\nu} \sum_X (\gamma_{AX} + \gamma_{BX}) \Delta q_X, \tag{4}$$

where the Hamiltonian matrix elements and the overlap matrix elements are pre-calculated as discussed above. Since the DFTB Hamiltonian depends explicitly on the atomic charge, a self-consistent charge (SCC) procedure is used in the SCC_DFTB approach to self consistently solve Eq. (4).

For the quantum dynamics calculations, the real-time, time-dependent DFTB (RT-TDDFTB) approach is utilized to propagate the one electron density matrix in the presence of external time-varying electric fields to obtain the time-dependent EET response of the system. This formalism has been previously used to probe the non-equilibrium electron dynamics in several large chemical systems,[25] including photoinjection dynamics in dye-sensitized $TiO_2$ solar cells[26,27] and many-body interactions in solvated nanodroplets.[28] We carry out our real-time quantum dynamics calculations by applying a time-dependent electric field to the initial ground state density matrix, resulting in the Hamiltonian $\hat{H}(t) = \hat{H}^0 - \mathbf{E}_0(t) \cdot \hat{\mu}(t)$, where $\mathbf{E}_0(t)$ is the applied electric field, and $\hat{\mu}$ is the dipole moment operator. As we are directly propagating the quantum system in the time-domain, we can choose to have any time-dependent form. For example, if $\mathbf{E}_0(t)$ is a Dirac delta function ($= \mathbf{E}_0 \delta(t - t_0)$) this corresponds to an optical absorption spectrum in the frequency domain (obtained after a Fourier transform of the time-evolving dipole moment). However if we choose $\mathbf{E}_0(t)$ to take the form of a sinusoidal perturbation, it represents a continuous interaction of the system with monochromatic light in the time domain. Both of these different choices give different but complementary viewpoints of quantum dynamics. Upon application of either of these time-dependent field, the density matrix, $\hat{\rho}$ will evolve according to the Liouville-von Neumann equation of motion which, in the nonorthogonal-DFTB basis, is given by

$$\frac{\partial \hat{\rho}}{\partial t} = \frac{1}{i\hbar}\left(\mathbf{S}^{-1} \cdot \hat{\mathbf{H}}[\hat{\rho}] \cdot \hat{\rho} - \hat{\rho} \cdot \hat{\mathbf{H}}[\hat{\rho}] \cdot \mathbf{S}^{-1}\right) \quad (5)$$

where $\hat{\mathbf{H}}$ is the Hamiltonian matrix (which implicitly depends on the density matrix), $\mathbf{S}^{-1}$ is the inverse of the overlap matrix, and $\hbar$ is Planck's constant. When the applied incident fields are smaller than the internal fields within the matter, the system is found to be in the linear response regime.[29] Under these conditions, the time evolution of the dipole moment operator can be expressed as the convolution between the applied electric field perturbation and the response function of the system:

$$\langle \hat{\mu}(t) \rangle = \int_0^\infty \alpha(t-\tau) E(\tau) d\tau, \quad (6)$$

where $E(\tau)$ is the electric field used to induce a perturbation in the system Hamiltonian, and $\alpha(t-\tau)$ is the polarizability tensor. Upon application of the convolution theorem, Eq. 6 can be expressed in the frequency ($\omega$) domain as $\langle \hat{\mu}(\omega) \rangle = \alpha(\omega) E(\omega)$. The imaginary part of the average polarizability, $\bar{\alpha}$ is an experimentally measurable quantity related to the photoabsorption cross section by the expression $\sigma(\omega) = 4\pi\omega/c \cdot \text{Im}(\bar{\alpha})$, where $c$ is the speed of light, and $\text{Im}(\bar{\alpha})$ is the imaginary part of the average polarizability. We utilized the DFTB+ code[38] to compute the ground-state Hamiltonian, overlap matrix elements, and the initial single-electron density matrix within the self-consistent DFTB approach. For the calculations performed here, we have used the matsci-0-3 set of Slater-Koster parameters for the sodium atoms.

**Local Surface Plasmonic Resonances in a Single Nanoparticle**

Before proceeding to a detailed analysis of the EET mechanism in plasmonic nanoantenna systems, we first characterize the LSPR of a single plasmonic NP. Accordingly, we plot the absorption spectrum of a single icosahedral shaped sodium NP (Na$_{55}$), containing 55 atoms and a diameter of 13 Å, using our RT-TDDFTB methodology. As shown in Figure 1(a), a prominent peak, which corresponds to the LSPR, is seen around 3.16 eV and is in agreement with previously published computational[39-40] and experimental results[41]. As stated earlier, LSPR excitations are associated with very large values of local field enhancements. Therefore, to unequivocally classify the observed excitation as plasmonic, we also plot the field enhancement around the $Na_{55}$ NP as shown in Figure 1(b). Specifically, the $Na_{55}$ NP is optically excited with an external sinusoidal electric field with its frequency equal to the plasmonic energy and polarized in the direction of its transition dipole moment. The electric field induced by plasmonic oscillations at any point in space is calculated/plotted using the following expression:

$$\boldsymbol{E}(\boldsymbol{r}) = \sum_i \frac{\Delta q_i}{4\pi\epsilon^0} \frac{(\boldsymbol{r}_i - \boldsymbol{r})}{||\boldsymbol{r}_i - \boldsymbol{r}||^3}, \quad (7)$$

and the enhancement, $\Gamma$ is calculated as follows:

$$\Gamma = \frac{|\boldsymbol{E}|^2(\omega)}{|\boldsymbol{E}_{appl}|^2(\omega)}, \quad (8)$$

where the applied field has the form $\boldsymbol{E}_{appl}(t) = \boldsymbol{E}_0 \sin(\omega t)$ in the time domain, and ω is the plasmon energy. As expected from plasmonic excitations, very high values of field enhancements are observed around the $Na_{55}$ NP, which are distributed in a dipolar fashion in alignment with the polarization vector **E**.

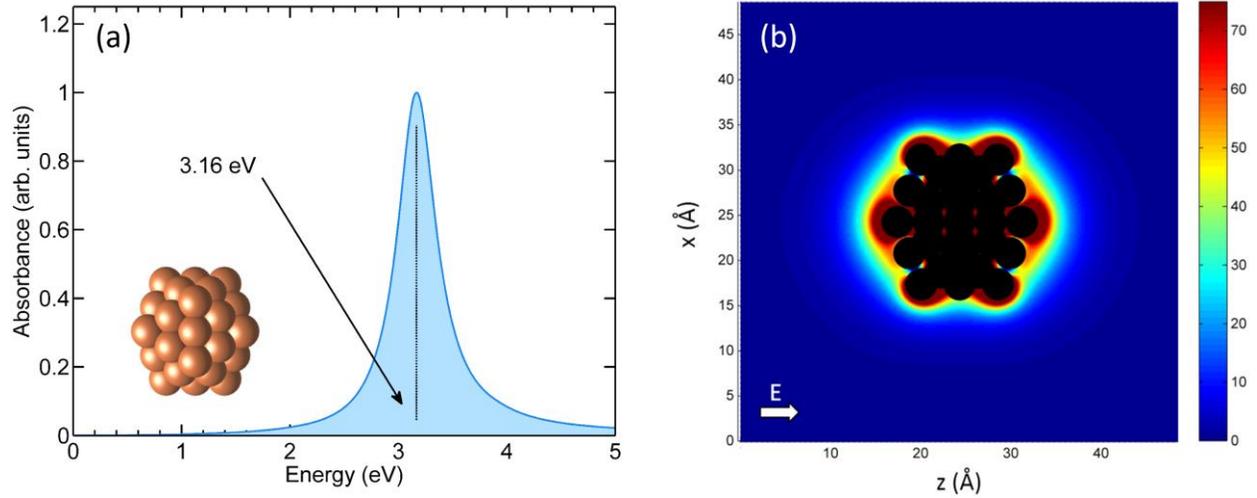

**Figure 1:** (a) Absorption spectra of the $Na_{55}$ NP (inset) calculated using RT-TDDFTB. The plasmon energy peak is observed at 3.16 eV; (b) Electric field enhancement of the $Na_{55}$ NP distributed in a dipolar fashion in alignment with polarization vector **E**. The dark spheres in (b) indicate the position of the Na atoms in the $Na_{55}$ NP.

We begin our investigation into the EET mechanism in plasmonic nanoantenna systems with an analysis of a simple donor-acceptor pair composed of two $Na_{55}$ NPs. The methodology used for calculating the time-dependent dipole moment due to a time-dependent electric field perturbation and the two-level system model (described later) is based on a previous study by one of the authors[42]. We placed a second identical $Na_{55}$ NP next to the first $Na_{55}$ at an edge-to-edge distance of 60 Å (= 73 Å center-to-center distance) with their transition dipole moments aligned with the *z*-direction, as shown in the inset of Figure 2. As previously mentioned, plasmon induced EET processes are analogous to FRET processes, where electronic excitation is transferred from a donor (NP1) to an acceptor (NP2) exclusively via electrostatic interaction. Thus, maintaining large interparticle distances (= 73 Å center-to-center distance) ensures that EET is attributed to purely electrostatic coupling between the donor and acceptor NPs. We also do not account for relativistic effects such as the retardation time of electromagnetic propagation between the NPs. Next; we excite only the donor (i.e. NP1) with a laser (sinusoidal electric field perturbation) with its frequency equal to the plasmon energy and polarized in the direction of the transition dipole moments (the z-direction). A very small intensity in the laser excitation ($E_0 = 0.0001$ V/Å) is used to ensure that we remain in the linear response regime[43]. We then allow the entire system to evolve in time for 50 fs and plot the induced dipole moments of both the NPs, as shown in Figure 2. Note that due to the orientation of the individual NPs and the polarization of the incoming radiation, the components of the dipole moment perpendicular to the NP chain (i.e., the x and y components) are essentially negligible compared to the z-component. We also note that the formalism does not include any dissipative mechanism and, hence, the dipole moments induced in the NPs are not damped with time. However, there have been a few recent studies to include these damping effects in the RT-TDDFT Hamiltonian,[44-45] and an equivalent approach could be applied to the RT-TDDFTB mechanism which we reserve for future work.

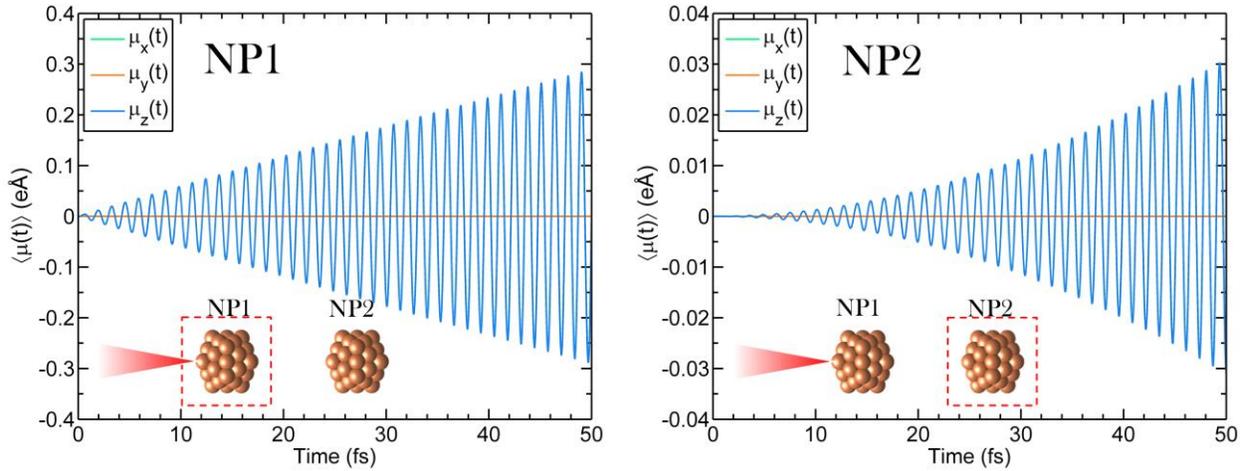

**Figure 2:** Time-dependent dipole moments induced in NP1 and NP2 nanoparticles upon optical excitation of NP1 with a sinusoidal electric field perturbation calculated using RT-TDDFTB. The induced dipole moment in NP2, is solely due to stimulation from NP1 and is indicative of electronic excitation transfer.

As shown in Figure 2, NP1 exhibits a linearly increasing dipole moment as expected from a quantized system in the linear response regime (in the presence of continuous excitation and in the absence of any dissipative mechanism)[42-43]. In contrast, NP2 displays an induced dipole moment that is entirely due to the stimulation provided by the oscillating electric field of NP1.

This induced dipole moment in NP2 is indicative of the EET process from the donor to the acceptor[46]. It is interesting to note that, even at such large separation distances (73 Å), strong dipole moments are induced in the second NP. We stress that this display of real-time EET between the acceptor-donor pair is obtained from a full quantum dynamical simulation without any approximations, such as the spectral overlap or dipole approximation for electronic couplings, typically considered in FRET based approaches[18]. The only approximations considered are the ones implied in the nature of the DFTB Hamiltonian[37]. Our methodology of exciting only the first NP in a closely spaced NP chain, though experimentally difficult, is based on previous computational studies and is meant to present an intuitive representation of the complex EET process.[47-49] We would like to point out that since we are exciting only the first NP, no collective excitations are observed in the system. However, the RT-TDDFTB calculations *do not* exclude the possibility of any energy back-transfer from the second to the first NP. If the system was allowed to evolve for a longer time period and with shorter interparticle distances, back-transfer of energy from the second to the first NP would be seen (see Supporting Information).

**A Two-Level System Model of Electronic Excitation Transfer**

While our RT-TDDFTB calculations fully incorporate electronic and atomistic details to characterize EET in this plasmonic donor-acceptor pair, to obtain deeper mechanistic insight into this complex quantum dynamical process, we formulate an analytical model based on a two level system (TLS) to highlight the basic physics that mediate interactions between the NP pair. As stated previously, this model is based on previous work carried out by one of the authors on EET mechanisms between photosynthetic pigments[42]. LSPR[3] is the coherent oscillation of electrons, between the ground and an excited state; therefore, the individual NPs can be approximated by a TLS, where the difference between the energy levels is equal to the plasmon energy. Additionally, since the size of the NPs is much smaller than the separation between them, the dipole approximation is justified; i.e., a point dipole interacting with another point dipole can be used to approximate the coupling between the NPs.

We derive a simple TLS model based on the above approximations to predict the induced dipole moment in NP2 as a result of the direct excitation of NP1 (Fig. 3). From linear response theory,[43] and considering each NP as a TLS, we can obtain closed-form analytical expressions of the expectation values of the dipole moment.

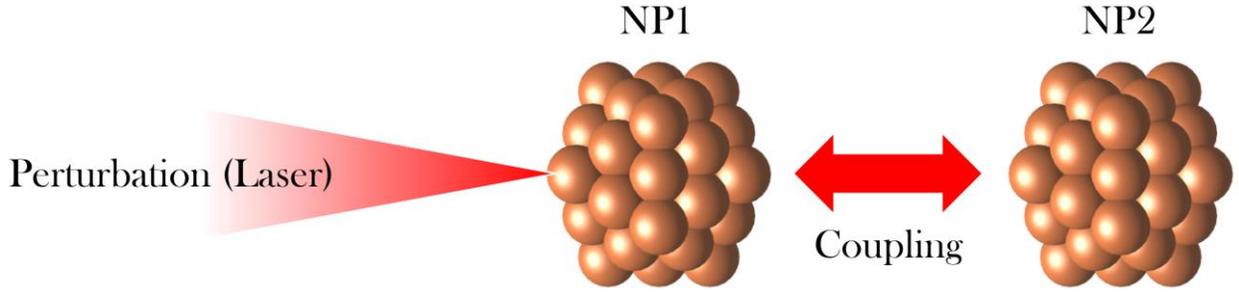

**Figure 3:** Pictorial representation of the TLS model for a Na$_{55}$ dimer. NP1 is optically excited with monochromatic light which induces a time-dependent dipole moment in NP2.

Within the linear response regime, the response of the dipole moment to a laser perturbation is given by

$$\boldsymbol{\mu}(t) = \int_0^\infty d\tau \, \mathbf{R}(\tau) \cdot \mathbf{E}(t-\tau), \tag{9}$$

where $\mathbf{E}(t-\tau)$ is the applied electric field and $\mathbf{R}(\tau)$ is the linear response function given by the expression

$$R_{\alpha,\beta}(\tau) = -\frac{i}{\hbar}\langle[\hat{\mu}_\alpha(\tau),\hat{\mu}_\beta]\rangle, \qquad \tau \geq 0 \tag{10}$$

$$= 0, \qquad \tau < 0$$

Here, $\langle[\hat{\mu}_\alpha(\tau), \hat{\mu}_\beta]\rangle$ is the polarizability tensor (expressed in terms of the commutator between $\hat{\mu}_\alpha(\tau)$ and $\hat{\mu}_\beta$) that describes the dipole moment response in direction $\alpha$ to an applied electric field in direction $\beta$. The linear response function is, therefore, the sum of two correlation functions with the order of the operators interchanged, which is further obtained from the imaginary part of the correlation function $C''(\tau)$:

$$R_{\alpha,\beta}(\tau) = -\frac{i}{\hbar}\{\langle\hat{\mu}_\alpha(\tau), \hat{\mu}_\beta\rangle - \langle\hat{\mu}_\beta, \hat{\mu}_\alpha(\tau)\rangle\}, \qquad (11)$$

$$= \frac{2}{\hbar} C''_{\alpha,\beta}(\tau).$$

Expectation values of observables are related to the imaginary part of the correlation function by the following definition:

$$C''_{\alpha,\beta}(\tau) = \frac{1}{2i}\left[\langle\hat{\mu}_\alpha(\tau), \hat{\mu}_\beta\rangle + \langle\hat{\mu}_\beta, \hat{\mu}_\alpha(\tau)\rangle\right]. \qquad (12)$$

Furthermore, the dipole moment observable can be represented in the interaction picture as

$$\boldsymbol{\mu}(\tau) = e^{-iE_{PE}\tau/\hbar} \cdot \boldsymbol{\mu}_{PE}, \qquad (13)$$

where $E_{PE}$ is the plasmon energy, and $\boldsymbol{\mu}_{PE}$ is the transition dipole moment. Substituting the above expressions into Eq. 10, we obtain the final expression for the response function given by

$$R_{\alpha,\beta}(\tau) = \frac{2}{\hbar}|\boldsymbol{\mu}_{PE}|^2 \sin(\omega_{PE}\tau)\, r_{PE}^\alpha r_{PE}^\beta, \qquad (14)$$

where $\omega_{PE} = \Delta E_{PE}/\hbar$ and $\hat{r}_{PE} = r_{PE}^x \hat{\imath} + r_{PE}^y \hat{\jmath} + r_{PE}^z \hat{k}$, such that the magnitude of $\hat{r}_{PE}$ is equal to 1.

As previously mentioned, to study the EET dynamics we apply a perturbation, $\mathbf{E}(t-\tau)$ in the form of a sinusoidal electric field given by

$$\mathbf{E}(t-\tau) = \mathbf{E_0} \sin[\omega_{PE}(t-\tau)]. \qquad (15)$$

Considering the applied field is in the direction of the transition dipole moment and substituting for the response function in Eq. 9 yields the time-evolving expectation value of the dipole moment in NP1:

$$\mu_\alpha(t) = \frac{2}{\hbar} E_0 |\boldsymbol{\mu}_1|^2 \int_0^\infty d\tau \sin(\omega_{\text{PE}}\tau) \sin[\omega_{\text{PE}}(t-\tau)] \, r_{\text{PE}}^\alpha \qquad (16)$$

As mentioned previously, the transition dipole moment of the NP and the applied electric field are both aligned in the *z*-direction. Hence, the solution of Eq. 16 for long times can be approximated by evaluating the integral and retaining only the term proportional to *t*:

$$\boldsymbol{\mu}_1(t) \approx \frac{E_0}{\hbar} |\boldsymbol{\mu}_1|^2 \, t \cdot \cos(\omega_{\text{PE}} t) \, \hat{r}_{\text{PE}} \qquad (17)$$

With the dipole moment in NP1 calculated, we next analyze the effect of this oscillating dipole on the acceptor NP (i.e., NP2). NP2 is located at a distance, $\hat{r}$, from NP1 where $|\hat{r}|$ is larger than the spatial extent of NP1 and NP2. The electric field generated by the oscillating dipole of NP1 at $\hat{r}$ is given by the following equation (i.e., the dipole approximation). We again clarify that we only invoke the dipole approximation in the TLS model since the electronic coupling between NPs at such large distances are well described by this approximation.

$$\mathbf{E}_1(t) = \frac{1}{4\pi\epsilon_0 r^3} \left( 3(\boldsymbol{\mu}_1(t) \cdot \hat{r})\hat{r} - \boldsymbol{\mu}_1(t) \right), \qquad (18)$$

where $\boldsymbol{\mu}_1(t)$ is the expectation value of the dipole moment of NP1 given by Eq. 17, $\epsilon_0$ is the vacuum permittivity, and $r$ is the distance between the nanoparticles. This oscillating electric field induces a dipole moment in NP2 which we obtain by following the methodology used previously for calculating the dipole moment in NP1. Particularly, we utilize Eq. 9 to calculate the expectation value of the dipole moment of NP2 due to the electric field induced by NP1:

$$\boldsymbol{\mu}_2(t) = -\frac{E_0}{4\pi\epsilon_0 \hbar^2 r^3} |\boldsymbol{\mu}_1|^2 |\boldsymbol{\mu}_2|^2 \sin(\omega_{\text{PE}} t) \, t^2 \left( \cos(\beta)\cos(\alpha) \right. \tag{19}$$

$$\left. - \frac{1}{2}\sin(\beta)\sin(\alpha) \right) \hat{r}_{\text{PE}},$$

where $\boldsymbol{\mu}_1$ and $\boldsymbol{\mu}_2$ are the transition dipole moments of NP1 and NP2, $\hat{r}_{\text{PE}}$ is the direction of the transition dipole moment of NP2, and $\alpha$ and $\beta$ are the angles between each dipole moment and the distance vector, $\hat{r}$. As mentioned previously, both nanoparticles are arranged with their transition dipole moments aligned along the *z*-direction. Furthermore, NP1 and NP2 are identical particles, and their transition dipole moments are equal to each other. Due to these simplifications, the above equation can finally be simplified to calculate the dipole moment induced in the *z*-direction as

$$\boldsymbol{\mu}_2(t) \approx -\frac{E_0}{4\pi\epsilon_0 \hbar^2 r^3} |\boldsymbol{\mu}_1|^4 \, t^2 \sin(\omega_{PE} t) \, \hat{r}_{PE}. \tag{20}$$

We denote Eqs. (17) and (20) as the TLS model. As previously mentioned, for the large interparticle distances and short time-periods considered, no back-transfer of energy is observed from the second to the first NP. Consequently, we ignore this back-transfer phenomena in our TLS approach and model the system as a unidirectional energy transfer system.

The value of the transition dipole moment for a single $Na_{55}$ NP is an initial condition parameter which we obtained from the RT-TDDFTB output of NP1 to Eq. 9. As mentioned previously, since both the NPs are identical to each other, this value is used to describe the transition dipole moments of both the NPs.

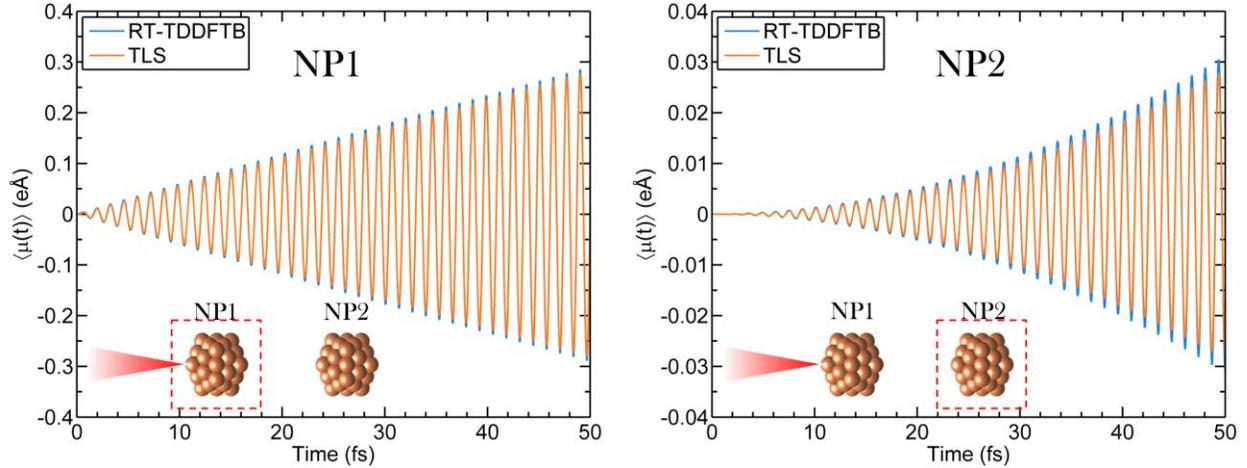

**Figure 4:** Comparison between the induced dipole moments calculated using the analytical two-level system (TLS) and RT-TDDFTB for NP1 and NP2.

Figure 4 compares the *z*-components of the dipole moments computed using the analytical TLS with the full quantum-dynamical RT-TDDFTB calculations. A close match between the two results suggests that our analytical model closely replicates the major results of the full quantum dynamical method. Specifically, we deduce that the long-range EET in plasmonic systems can be accurately described within the dipole approximation; that is, as long as the interparticle distance is relatively larger than the NP cross-section, plasmonic NPs can be treated as interacting point dipoles. While similar results have been analyzed in previous studies[10, 15, 17, 50], the RT-TDDFTB approach allows a fully electronic/atomistic treatment of these systems without recourse to any of the approximations made by the aforementioned studies.

With the EET in the simple plasmonic donor-acceptor pair fully characterized, we next turn our attention to a multi-particle plasmonic nanoantenna composed of 4 identical $Na_{55}$ NPs as shown in Figure 5. The inter-particle distance (center to center) is again set to 73 Å, and each NP is oriented with their transition dipole moment aligned in the *z*-direction. As before, we excite only NP1 using a laser with its energy tuned to the single $Na_{55}$ NP plasmon energy and polarized in the *z*-direction. Figure 5 plots the time-dependent dipole moments induced in all of the NPs within the nanoantenna system. The RT-TDDFTB calculations predict substantial dipole moments being induced in all of the NPs, indicating EET from the excited NP1 to the remaining NPs along the nanoantenna, corroborating previous experimental observations of EET observed in a chain of metallic NPs[8]. We note that due to the large distances involved (73 Å) and in the small time period considered (50 fs), the EET is unidirectional (NP1 to NP4) in nature; i.e., EET back-transfer is negligible.

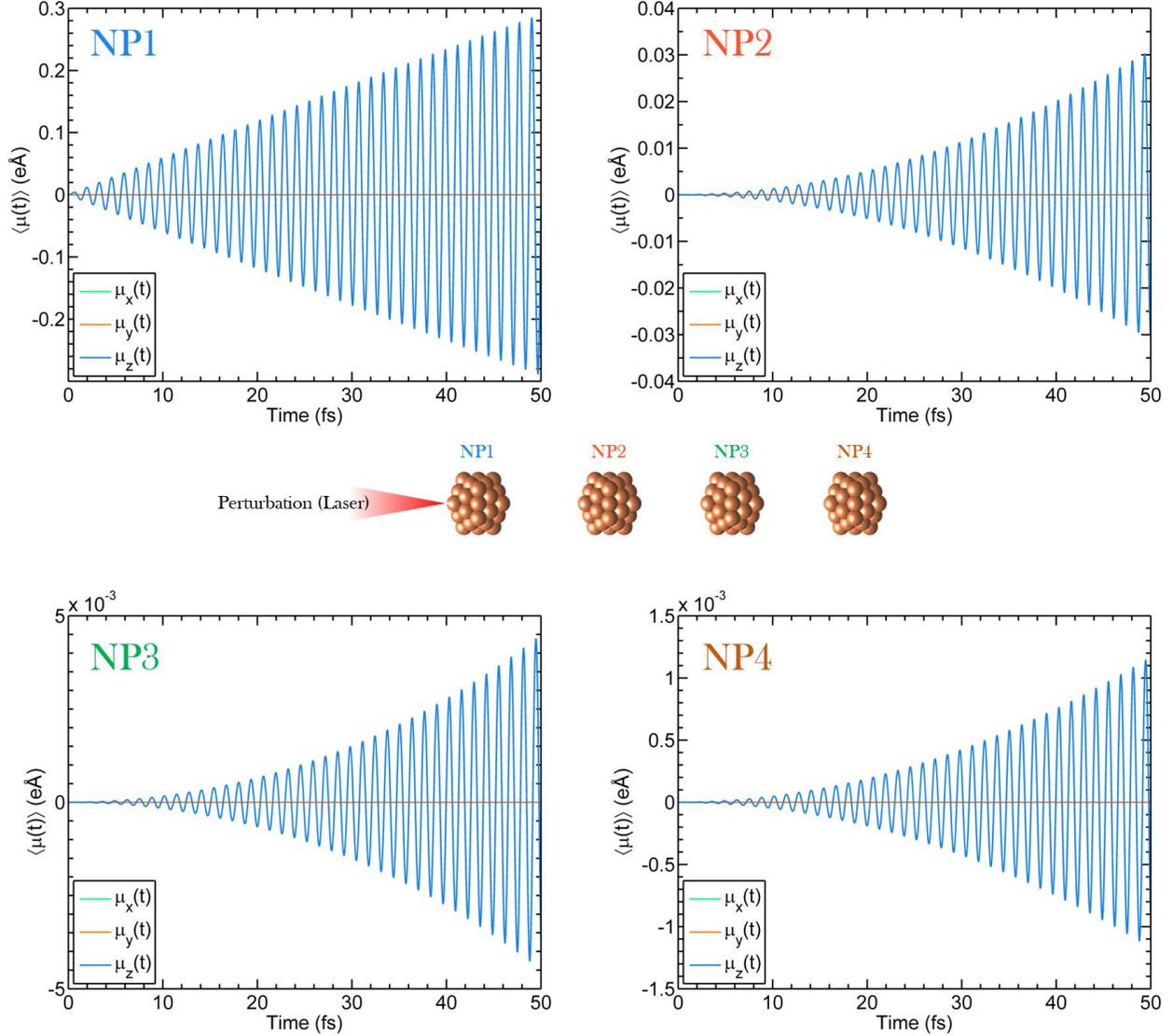

**Figure 5:** Time-dependent dipole moments induced in the four NPs of the plasmonic nanoantenna system upon optical excitation of NP1 with a sinusoidal electric field perturbation calculated using RT-TDDFTB. The induced dipole moments in the NPs are indicative of the electronic excitation transfer in the multi-particle plasmonic nanosystem.

Next, we extend our analytical TLS model to highlight the basic physics of the EET mechanism in this multi-particle plasmonic nanosystem. However, to expand the TLS model to capture the EET dynamics beyond the second NP, we first approximate the interactions between the various NPs to be limited to only nearest neighbor interactions. For instance, NP2 is only stimulated via oscillations in NP1, NP3 only due to NP2, and so on. This approximation is a commonly used assumption used in many classical electrodynamic approaches[50-52] and is also based on the maximum cutoff distance (i.e., 10 nm), considered by FRET approaches[13, 53], beyond which the EET is considered negligible. Specifically, in our plasmonic nanoantenna system, all the NPs, except for the nearest neighbors lie well beyond this cutoff distance. Furthermore, we utilize the dipole approximation, which we had shown in the previous section to accurately describe the individual interactions between each NP pair.

**Two-Level System Model for Four NP System (Including only Nearest-Neighbor Interactions)**

We expand the two-particle analytical TLS model to describe the four-particle plasmonic nanoantenna system depicted below in Figure 6.

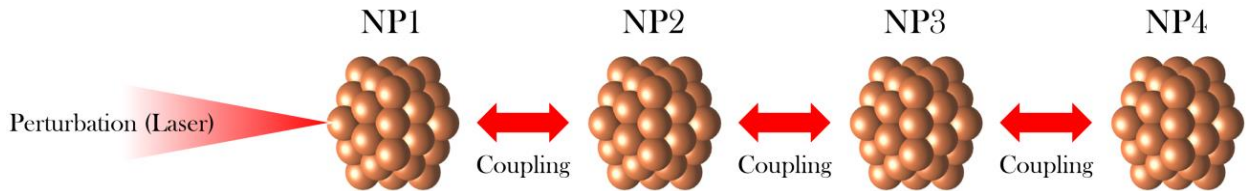

**Figure 6:** Pictorial representation of the TLS model for a Na$_{55}$ tetramer. The first NP is optically excited with monochromatic light which induces a time-dependent dipole moment in the following NPs. The arrows represent the couplings considered (i.e., nearest-neighbor) in the TLS model for the plasmonic nanoantenna.

We utilize the previous described theoretical approach to derive the induced dipole moment in NP3. Also, as previously mentioned, we use the nearest-neighbor interaction model in which only interactions between nearest neighbors are considered for EET. Therefore, the dipole moment induced in the *z*-direction in NP3 is only due to the oscillating electric field of NP2, and the expectation value of the dipole moment in NP3 is given by

$$\boldsymbol{\mu}_3(t) \approx \frac{E_0}{24\pi^2 \epsilon_0^2 \hbar^3 r^6} |\boldsymbol{\mu}_1|^6 \, t^3 \cos(\omega_{\text{PE}} t) \, \hat{r}_{\text{PE}}. \tag{21}$$

Similarly, the dipole moment in *z*-direction for NP4, considering its interaction only with NP3, is given by

$$\boldsymbol{\mu}_4(t) \approx \frac{E_0}{192\pi^3 \epsilon_0^3 \hbar^4 r^9} |\boldsymbol{\mu}_1|^8 \, t^4 \sin(\omega_{\text{PE}} t) \, \hat{r}_{\text{PE}}. \tag{22}$$

Figure 7 compares the dipole moments calculated using the expanded TLS model with the RT-TDDFTB results. While the results of our analytical TLS model match closely with the RT-TDDFTB results for NP1 and NP2, it grossly underestimates the dipole oscillations in NP3 and NP4. Since we have categorically examined the validity of the dipole approximation in the previous section, the failure of the analytical model indicates that the "nearest-neighbor" approximation considered in the multi-particle model is the culprit. The validity of this "nearest-neighbor" approximation has also been previously contested by Citrin[17] and co-workers.

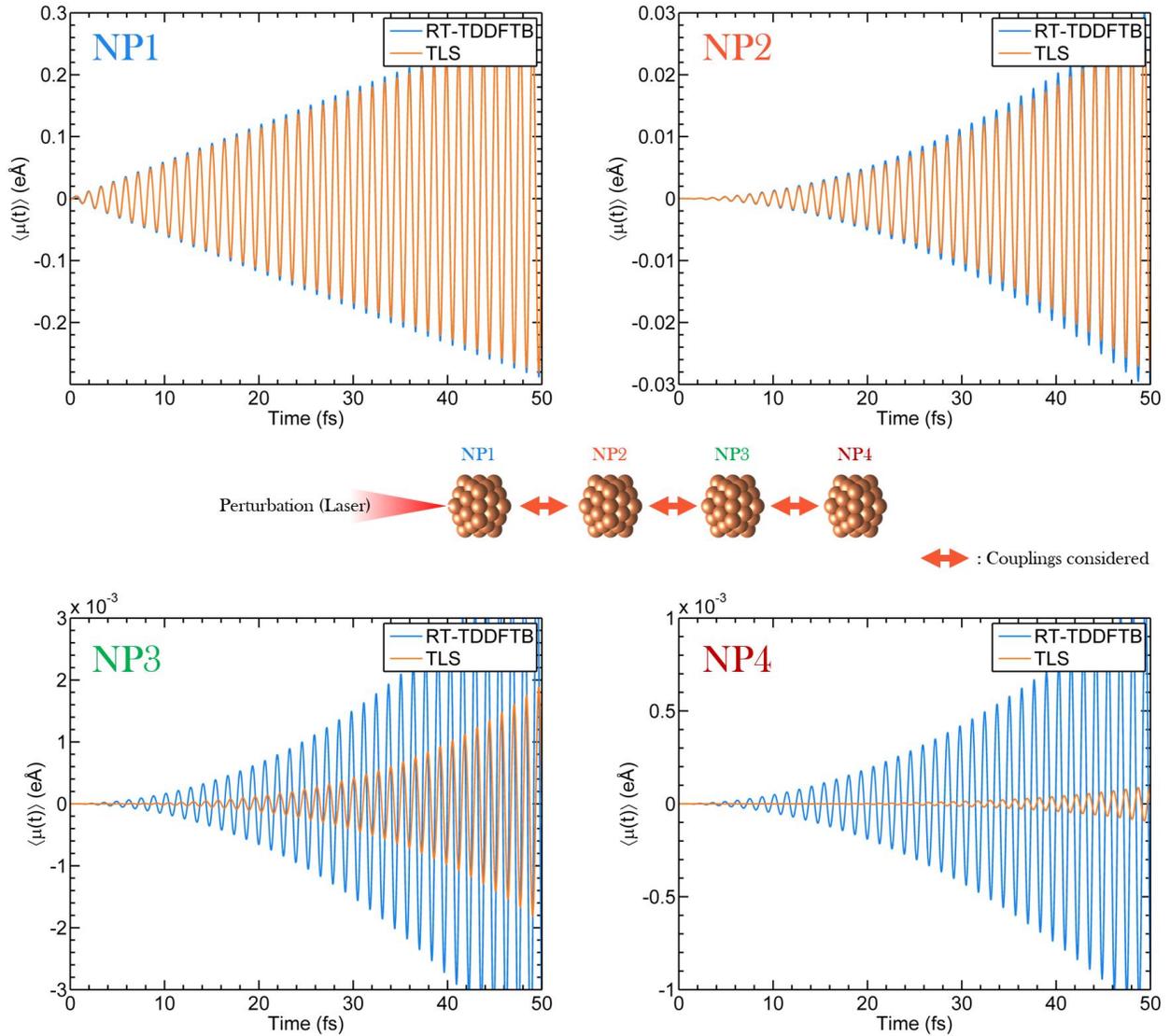

**Figure 7:** Comparison between the dipole moments calculated using the analytical two-level system model (TLS) that considers only the nearest-neighbor interactions and RT-TDDFTB calculations for the plasmonic nanoantenna. The dipole moments in NP3 and NP4 are severely underestimated by this analytical model.

To prove this conjecture, we modify the TLS model to include interactions between all the NPs in the entire nanoantenna. For example, NP4 is stimulated collectively by NP1, NP2 and NP3, and so on; however, we still use the dipolar approximation to describe the individual NP interactions.

**Modified Two-Level System Model Including all Interactions**

The primary modification in this version of the analytical model is that we now consider *all* of the inter-particle interactions. For instance, the dipole moment in NP3 arises from its interaction with NP2 and NP1 (only the interaction with NP2 was considered in the previous nearest-neighbor TLS model). Similarly, the dipole moment in NP4 is due to its interaction with all the other NPs; i.e., NP1, NP2, and NP3. This modified scheme with all of the couplings is shown in Figure 8.

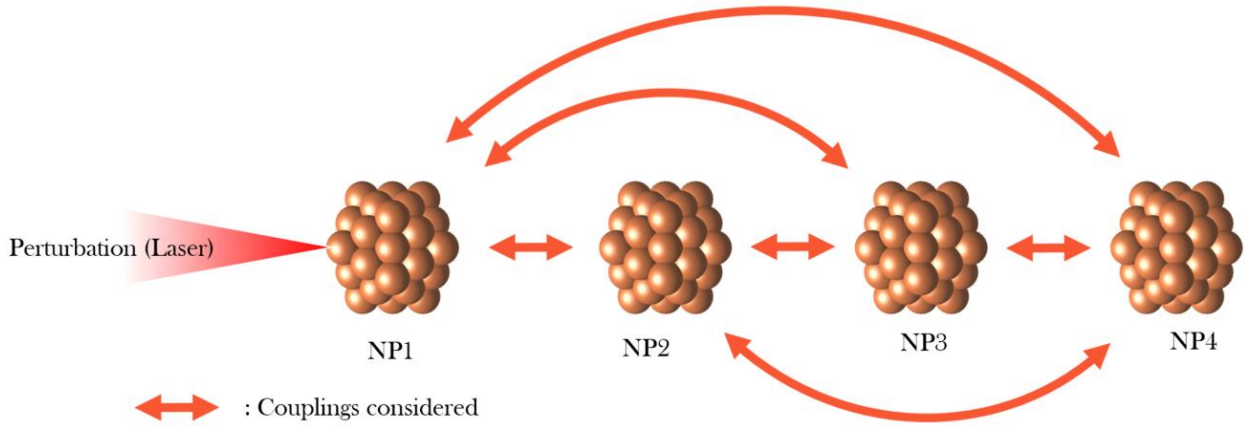

**Figure 8:** Pictorial representation of the TLS model for a Na$_{55}$ tetramer. The first NP is optically excited with monochromatic light which induces a time-dependent dipole moment in the following NPs. The arrows represent the couplings considered (i.e., all interactions) in the TLS model for the plasmonic nanoantenna.

Using the same approach described previously, the new equations for calculating the dipole moments in NP3 and NP4 (in *z*-direction) are

$$\boldsymbol{\mu}_3(t) \approx \frac{E_0}{24\pi^2\epsilon_0^2\hbar^3 r^6}|\boldsymbol{\mu}_1|^6 t^3 \cos(\omega_{\text{PE}}t)\,\hat{r}_{\text{PE}} \tag{23}$$

$$-\frac{E_0}{4\pi\epsilon_0\hbar^2(2r)^3}|\boldsymbol{\mu}_1|^4\, t^2 \sin(\omega_{\text{PE}}t)\,\hat{r}_{\text{PE}},$$

and

$$\boldsymbol{\mu}_4(t) \approx \frac{E_0}{192\pi^3\epsilon_0^3\hbar^4 r^9}|\boldsymbol{\mu}_1|^8\, t^4 \sin(\omega_{\text{PE}}t)\, \hat{r}_{\text{PE}} \tag{24}$$

$$+ \frac{E_0}{24\pi^2\epsilon_0^2\hbar^3(2r)^6}|\boldsymbol{\mu}_1|^6\, t^3 \cos(\omega_{\text{PE}}t)\, \hat{r}_{\text{PE}}$$

$$+ \frac{E_0}{24\pi^2\epsilon_0^2\hbar^3(2r)^3}|\boldsymbol{\mu}_1|^6\, t^3 \cos(\omega_{\text{PE}}t)\, \hat{r}_{\text{PE}}$$

$$- \frac{E_0}{4\pi\epsilon_0\hbar^2(3r)^3}|\boldsymbol{\mu}_1|^4 t^2 \sin(\omega_{\text{PE}}t)\, \hat{r}_{\text{PE}},$$

respectively.

The induced dipole moments predicted by the new analytical model closely match the RT-TDDFTB results and are summarized in Figure 9. This modified TLS model illuminates a few more significant features of the EET mechanism in plasmonic nanoantennas. Most importantly, we note that the range of electronic couplings in plasmonic nanosystems is much larger than the FRET-based cutoff limits, and restricting couplings to the conventional FRET limit severely underestimates the EET in the plasmonic nanoantenna. For instance, when couplings only within the FRET limit are considered, the predicted EET in NP4, as shown in Figure 7, is an order of magnitude lower than the true EET predicted by the RT-TDDFTB calculations. Furthermore, as elucidated from the analytical model, the commonly used nearest-neighbor interaction model falls short in accurately predicting EET in plasmonic nanoantennas. A more complete multi-particle interaction model, which considers interactions between all the NPs of the nanoantenna, is needed to fully characterize such a system. Finally, we advise caution on the direct use of single donor-acceptor based approaches[22, 24], to model even simple multi-particle plasmonic systems, such as the one considered in this study. Comparisons between our full RT-TDDFB calculations with the simplified nearest-neighbor analytical models emphasize the severe limitations in single donor-acceptor models for accurately describing large multi-particle systems.

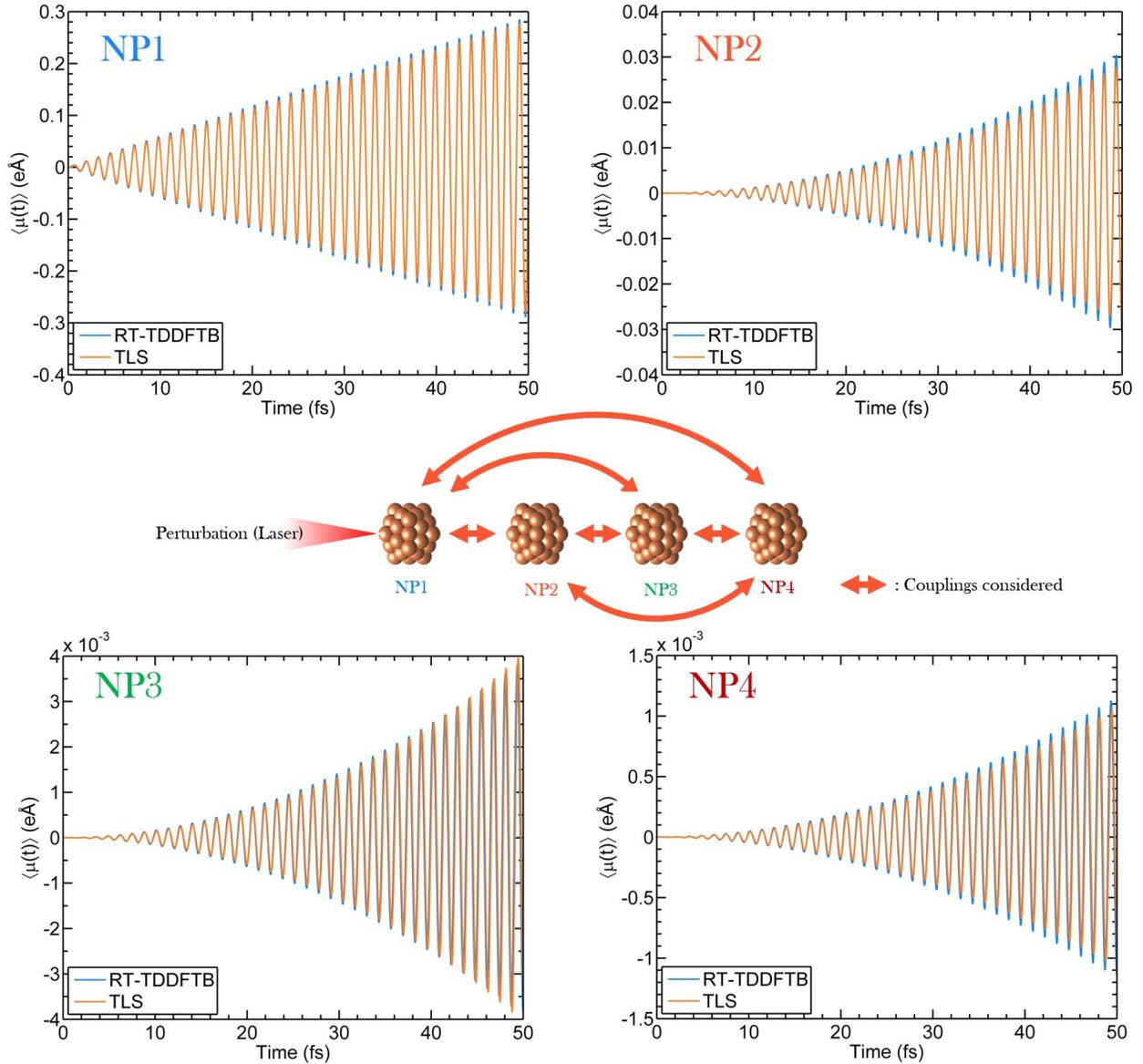

**Figure 9:** Comparison between the dipole moments calculated using the analytical two-level system (TLS) model, which considers interactions between all the particles and RT-TDDFTB calculations for the plasmonic nanoantenna. The multi-particle analytical model, that includes the long-range interactions, is accurately able to corroborate the RT-TDDFTB results.

Furthermore, in order to analyze the long-range plasmonic interactions in greater detail, we decomposed the RT-TDDFTB results for the total excitation induced in each of the NP into individual contributions due to the other NPs. For instance, the total electronic excitation in NP4 is decomposed into dipole moments due to individual stimulations from NP3, NP2, and NP1.

Figure 10 summarizes the total dipole moment induced in NP3 and NP4 as a combination of dipole moments due to stimulation provided by the other NPs. Note that NP1 and NP2 do not have such a decomposition since NP2 is stimulated solely due to NP1 as the laser directly excites NP1.

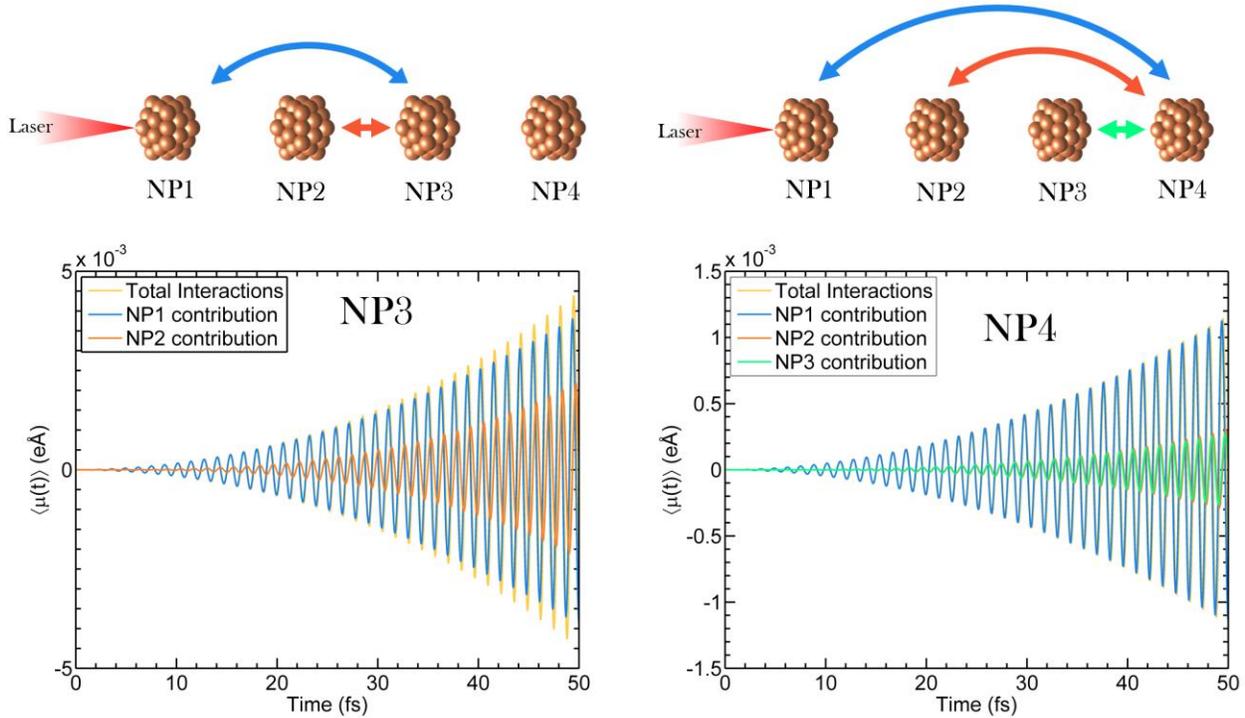

**Figure 10:** Total dipole moments induced (a) NP3 and (b) NP4 of the plasmonic nanoantenna decomposed into contributions by the other NPs in the system. Direct and substantial electronic excitation transfer is observed even between the farthest NPs.

Figure 10 re-emphasizes the long-range of the plasmonic EET that we had noted previously while developing the analytical model for the multi-particle plasmonic system. In particular, we observe that there is a direct and rather substantial EET even between the farthermost NPs. For instance, direct and substantial EET is seen between NP1 and NP4, despite the fact that the interparticle distance between these two NPs is ~ 21 nm, which is more than twice the conventional FRET cutoff distance. The presence of EET over such exceedingly long distances points towards the long-range nature of electrostatic couplings in plasmonic systems. We attribute this long range of electronic coupling, to the coherent nature of the plasmon resonances, where a large number of conduction electrons oscillate simultaneously to produce a large dipole moment. On account of this larger dipole moment, the electronic couplings of plasmonic systems extend well beyond the generally accepted FRET maximum cutoff limit for EET processes. We deduce that these conventional FRET cutoff limits are unsuitable for plasmon-induced EET because these cutoff limits were originally based on a single electron oscillating between the excited and ground state, thereby reducing the amplitude of the dipole moment and hence the coupling distance. Our prediction of these extremely long-range plasmonic couplings is also supported by previous experimental observations[13], where EET was detected in plasmonic rulers composed of gold NPs separated by more than 20 nm.

Returning to Figure 10, a closer look at the results also implies that each of the NPs in the nanoantenna system works simultaneously as a donor and a receiver. For instance, in Figure 10(a), NP3 while behaving as an acceptor for EET from NP1 and NP2 simultaneously behaves as a donor to NP4 (as indicated by its EET contribution to NP4 in Figure 10(b)). To highlight the uniqueness of this mechanism for plasmonic systems, we compare these results to a similar nanoantenna composed of *non-plasmonic* coronene nanoflakes.

**Non-Plasmonic Nanoantenna Composed of Coronene Nanoflakes**

To emphasize the long-range nature of *plasmonic* interactions, we construct a non-plasmonic antenna composed of coronene nanoflakes, as shown in Figure 11.

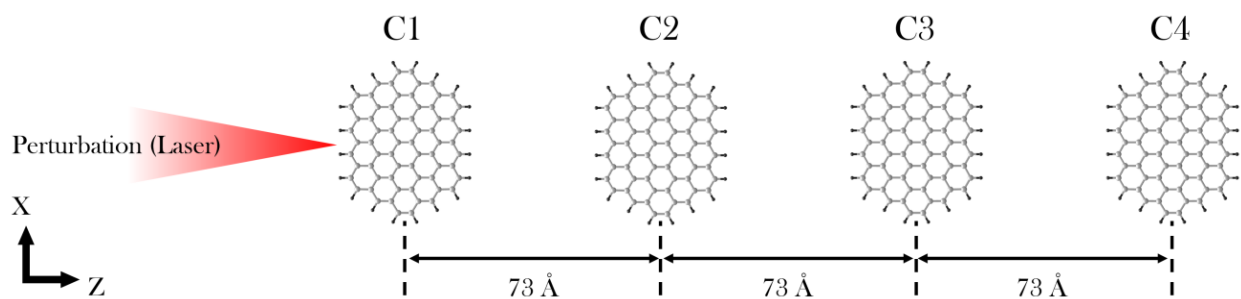

**Figure 11:** Pictorial representation of a non-plasmonic nanoantenna composed of four coronene flakes. The first coronene molecule is optically excited with monochromatic light which induces a time-dependent dipole moment in the following coronene molecules.

To maintain a fair and consistent comparison between the two systems, the nanoflakes were constructed with a diameter equal to the plasmonic NPs used previously (~13 Å) and aligned in the same spatial arrangement as the plasmonic nanoantenna. Analogous to the plasmonic case, only the first coronene flake was excited with a laser tuned to the first excitation peak (1.86 eV) observed in the coronene flake absorption spectrum. The dipole moments induced in each of the coronene flakes, calculated using the RT-TDDFTB calculations and decomposed into their contributions, are shown in Figure 12.

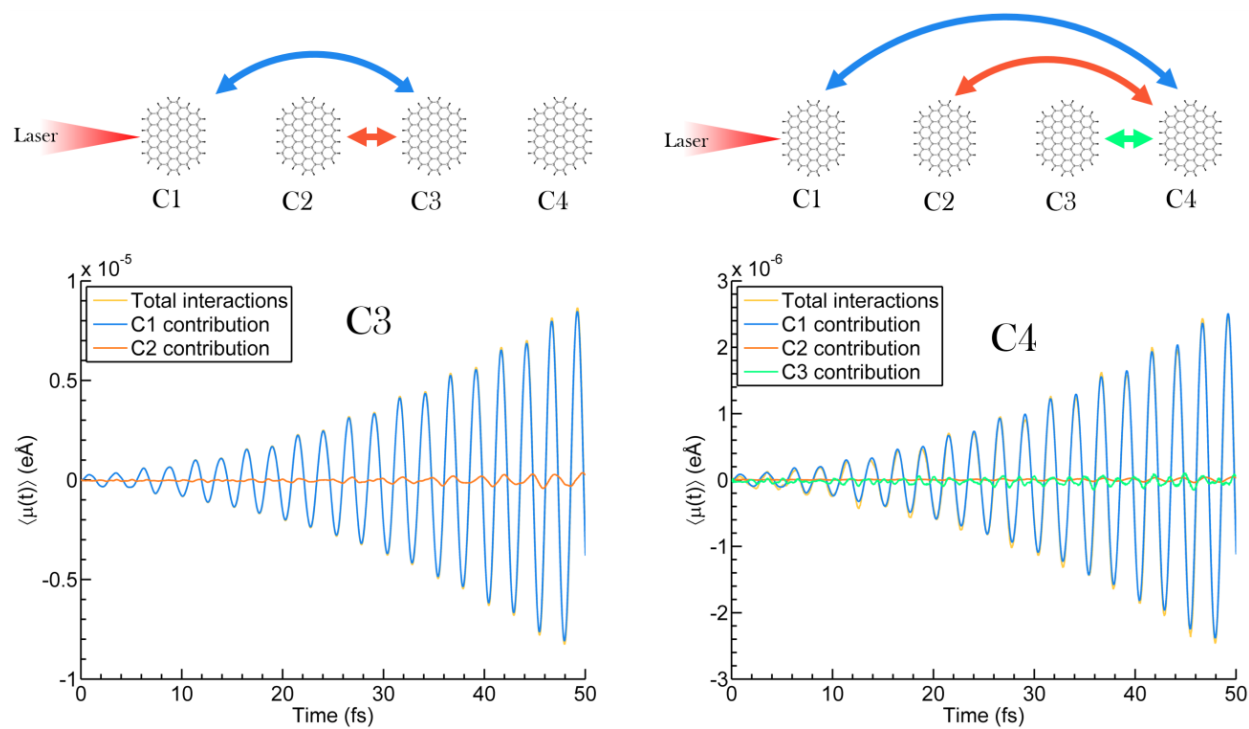

**Figure 12:** Total dipole moments induced in (a) C3 and (b) C4 of the non-plasmonic nanoantenna decomposed into contributions due to the other coronene nanoflakes in the system.

We observe that, unlike in the plasmonic case, the excitation reaching any nanoflake is almost entirely due to the oscillating electric field of the directly excited nanoflake (C1). For example, the EET in C4 is entirely contributed by C1, with the other nanoflakes in the nanoantenna contributing minimal excitation. Specifically, the directly-excited coronene flake acts like a dipole antenna, with all of the other coronene flakes acting as receivers. As a result, the EET mechanism in non-plasmonic antennas can be simplified to a single-donor, multiple-acceptor system and is much closer to the FRET mechanism. Therefore, contrary to the plasmonic case, in non-plasmonic antennas, the excited nanoflake works as the sole donor with the other nanoflakes behaving as acceptors. Thus, the EET mechanism in plasmonic antennas is unique and involves a multitude of multiple-donor, multiple-acceptor interactions that go beyond the single donor-acceptor mechanism found in non-plasmonic systems. It is also worth mentioning that even though EET occurs in non-plasmonic nanoantennas, the amplitude of the induced dipole moment is *several* orders of magnitude lower than the comparable plasmonic case. For example, the magnitude of the dipole moment in nanoflake C4 is at least three orders of magnitude lower than the dipole moment in NP4. This highlights the effectiveness of plasmonic systems for long-range energy transfer compared to an organic / non-plasmonic system.

**Conclusion**

In summary, we have thoroughly characterized the EET mechanism in a representative plasmonic nanoantenna system using large-scale RT-TDDFTB calculations that are further rationalized by various analytical two-level model systems. Most importantly, the RT-TDDFTB simulations provide a natural approach to probe in atomistic detail the time-dependent electron dynamics in multibody plasmonic systems without recourse to customary approximations, such as nearest-neighbor, spectral overlap, or the dipole approximations to describe electronic couplings. Furthermore, we reveal highly long-range plasmonic couplings that are more than twice the conventional cutoff limit considered by FRET based approaches. We attribute this unusually higher range of electronic couplings to the coherent oscillation of conduction electrons in the plasmonic NPs. Due to the collective nature of the oscillating electrons, the magnitude of the dipole moment produced is substantially larger than the dipole moment of a single oscillating electron, typically considered in FRET approaches, thereby increasing the range of plasmonic interactions. An important ramification of this long-range nature of plasmonic EET is that the "nearest-neighbor" interaction model commonly used to characterize EET is highly inadequate for plasmonic systems, even in unidirectional plasmonic antennas such as the one considered in this study. A more complete model, which considers interactions between all of the constituents in the nanoantenna system, is therefore needed to correctly determine the EET processes. These analytical models both complement and corroborate the RT-TDDFTB calculations to highlight mechanistic details that go beyond nearest-neighbor approaches for plasmonic nanoantennas. While the use of short-ranged FRET-based approaches have long been used to characterize plasmonic systems, our findings strongly emphasize the importance of long-range, multiple-particle interactions in mediating the EET dynamics of these systems. Consequently, our results provide a new viewpoint for characterizing and understanding these systems for harnessing and controlling long-range transfer of excitation energy in increasingly complex plasmonic nanosytems.

## ASSOCIATED CONTENT

## Acknowledgments

This work was supported by the U.S. Department of Energy, Office of Science, Early Career Research Program under Award No. DE-SC0016269. We acknowledge the National Science Foundation for the use of supercomputing resources through the Extreme Science and Engineering Discovery Environment (XSEDE), Project No. TG- ENG160024.